\documentclass[12pt]{iopart}
\usepackage{graphicx}
\usepackage{iopams}
\usepackage[alsoload=synchem]{siunitx}
\usepackage{braket}
\usepackage{hyperref}
\usepackage{epstopdf}
\usepackage{url}
\usepackage{multirow}
\usepackage{dcolumn}
\usepackage{bm}
\begin{document}

\title[Light-shift spectroscopy of optically trapped atomic ensembles]{Light-shift spectroscopy of optically trapped atomic ensembles}

\author{Ashby P Hilton$^1$, Andre N Luiten$^1$, Philip S Light$^1$}
\address{$^1$ Institute for Photonics and Advanced Sensing (IPAS) \& School of Physical Sciences, The University of Adelaide, Adelaide, South Australia 5005, Australia}
\ead{Ashby.Hilton@Adelaide.edu.au}
\vspace{10pt}
\begin{indented}
	\item[]October 2019
\end{indented}

\begin{abstract}
	
	We develop a method for extracting the physical parameters of interest for a dipole trapped cold atomic ensemble.
	This technique uses the spatially dependent ac-Stark shift of the trap itself to project the atomic distribution onto a light-shift broadened transmission spectrum.
	We develop a model that connects the atomic distribution with the expected transmission spectrum.
	We then demonstrate the utility of the technique by deriving the temperature, trap depth, lifetime, and trapped atom number from data that was taken in a single shot experimental measurement.
	
\end{abstract}


\section{Introduction}
	\label{Section: Introduction}
	
	Cold atomic ensembles are a staple in the world of precision measurement and fundamental physics due to the high atomic densities and low kinetic energies attainable.
	These properties, combined with a wide range of internal degrees of freedom, make for a versatile tool capable of high sensitivity measurement \cite{Kasevich1991,Peters2001,Canuel2006,Cronin2009,Stockton2011,Hinkley2013,Altin2013,Dutta2016}, quantum storage and manipulation \cite{Kasevich1991,Sparkes2013,Liu2016,DeAlmeida2016,Park2018,Hsiao2018}, and highly accurate metrology \cite{Katori2003,Takamoto2005,Ye2008,Bloom2014,Huang2014,Zhang2016}.
	
	In many cases optical dipole traps are used to localise the atomic ensemble in a controllable way.
	This can be used to hold the atomic sample for a long duration \cite{Takamoto2005,LeTargat2013,Gross2017} or to transport the atoms into a confined geometry such as a hollow-core fibre \cite{Vorrath2010,Bajcsy2011,Okaba2014,Blatt2016,Xin2017,Hilton2018,Langbecker2018,Yoon2019} or near to a structured device such as an atomic chip \cite{Chikkatur2002,Leanhardt2002a,Leanhardt2002b,Leanhardt2003}.	
	The act of trapping can itself alter the temperature and size of the ensemble from that which might be derived from a post-facto measurement.
	
	A standard technique for calculating temperature of a trapped ensemble is direct absorption imaging, by which the ensemble is released from the trap and illuminated with resonant light, casting a silhouette or shadow onto a waiting camera \cite{Ketterle1999}.
	While this is effective, it is necessary to take many images at various times after release to obtain reliable temperature information.
	As the imaging pulse imparts significant momentum on the atoms, each time slice must be taken on a separate run of the experiment, making this a slow process that can be susceptible to experimental fluctuations over multiple imaging cycles.
	Additionally, this method requires direct optical access to both sides of the atomic cloud, which is not always possible.
	
	A similar approach is the release and recapture technique, which also uses the ballistic expansion of the cloud to estimate its thermodynamic properties.
	This technique is commonly used to measure the temperature of cold-atom ensembles confined to hollow-core fibre, as the atoms can be interrogated using a weak probe field overlapping the trapping beam \cite{Peyronel2012,Blatt2014,Xin2017,Langbecker2018,Yoon2019}.
	However, multiple interrogation sequences with increasing free expansion periods are still necessitated to extract the temperature of the ensemble.
	
	Our approach is an in-situ method that exploits the effect of the trapping field's ac-Stark shift on the absorption of the trapped atoms.
	This can be used to map the location of an atom in the trap into a shift of the atomic transition through the spatially varying trap intensity.
	We calculate the expected atomic distribution within a stable Gaussian trapping field, and identify how temperature and other properties of the ensemble affect the functional form of the light shifted absorption spectrum.
	We show that each of the interesting parameters are sufficiently different in their effect on the absorption spectrum that they can be individually extracted from a single measurement of the broadened spectrum.
	This negates the need for destructive release and recapture techniques.
	
	We also consider the sensitivity of this technique to the shape of the atomic distribution.
	It has been suggested within the community that the spatial distribution of atoms within a trap might be ring-like, with its peak density away from the central axis.
	We develop a self-consistent model for the atomic distribution that produces such an atomic profile and apply our light-shift spectroscopy approach to calculate the absorption spectrum that would be obtained under this assumption.
	We test both the conventional Gaussian model and the ring-like model against experimental data gathered from our experiment and that of Peyronel et al. \cite{Peyronel2012}, and show that our ring-like model does not well match the experimental results.
	Using the Gaussian model, we find a good agreement with the measured spectra, and are able extract a range of experimentally useful parameters.
	This demonstrates the power of this technique for performing fast, single-shot interrogation of dipole-trapped atomic ensembles.
	This approach also has great merit under some circumstances, such as when direct imaging is not feasible.
	In addition, as it is not necessary to modulate or switch off the trapping field during measurement, this process is relatively non-destructive allowing multiple measurements to be made during a single experimental run.
	
\section{Light-shift spectroscopy}
	\label{Section: Light-Shift Spectroscopy}
	
	The basis of optical dipole trapping is well known: a strong light field far red-detuned from a two-level atomic system produces a reduction in energy of the ground state.
	This perturbation is proportional to the local intensity of the dipole beam, and as such a typical Gaussian beam can produce a potential capable of confining atoms.
	The spatially dependent energy shift serves not only to trap the atoms, but also produces a shift in the line-centre of any transition from the ground state, typically referred to as a light-shift.
	This effect is typically undesirable, producing an additional source of ensemble decoherence and broadening of the transition \cite{Haffner2003,Hong2005,Santra2005}.
	However this lifting of the spatial degeneracy of the atomic ensemble can be performed in a predictable way, giving rise to the possibility of extracting information regarding the distribution of atoms in the trapping light field and the absolute depth of the trap.
	To understand the effect of both the trap profile and atomic distribution on the measurable absorption spectrum, we build a model using a typical single beam optical trap and a thermal atomic ensemble.
	
	\begin{figure}
		\begin{indented}
			\item[]
			\includegraphics{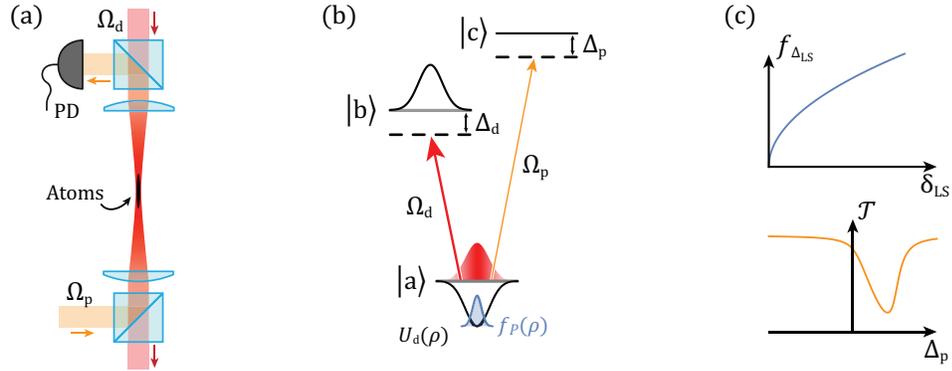}
			\caption{
				\label{Figure: Concept Plot}
				a) The basic experiment system typically used, in which the probe field $\left(\Omega_\mathrm{p}\right)$ is combined with and separated from the trap field  $\left(\Omega_\mathrm{d}\right)$, and incident on a photo-detector (PD).
				b) The energy level structure modelled with the ground-state $\ket{\mathrm{a}}$, upper trap-state $\ket{\mathrm{b}}$, and upper probe-state $\ket{\mathrm{c}}$, including the perturbative effect of the dipole trap, $U_\mathrm{d}\left(\rho\right)$, and the resulting radial probability distribution function $f_{P}\left(\rho\right)$.
				c) The calculated atomic distribution as a function of the light-shift (top), and the resulting in-trap transmission spectrum as measured by the probe (bottom).
			}
		\end{indented}
	\end{figure}
	We consider the potential generated by a collimated dipole beam far detuned from an atomic transition, which can be written in cylindrical coordinates as 
	\begin{equation}
		\label{Equation: Dipole Trap}
		U_\mathrm{d}\left(\bi{r}\right)=U_\mathrm{d}\left(\rho,\theta,z\right)=U_0 u\left(\rho\right)
	\end{equation}
	where $U_0$ is the peak trap amplitude, and $u\left(\rho\right)$ is the relative optical intensity in the radial direction $\hat{\brho}$.
	For a Gaussian dipole beam with power $P_\mathrm{d}$ and waist $w$, the relative intensity is simply $u\left(\rho\right)=\exp(-2\rho^2/w^2)$, and the depth of the trap on axis is
	\begin{equation}
		\label{Equation: Trap Depth}
		U_0=\frac{3\pi c^2}{2 \omega_0^3} \frac{\Gamma_\mathrm{d}}{\Delta_\mathrm{d}} \frac{2P_\mathrm{d}}{\pi w^2}
	\end{equation}
	where $c$ is the speed of light, $\omega_0$ and $\Gamma_\mathrm{d}$ are the angular frequency and natural linewidth of the atomic transition, and $\Delta_\mathrm{d}$ is the angular frequency detuning of the dipole trap laser from $\omega_0$.
	
	In the thermal regime the atomic density $n\left(\bi{r}\right)$ is determined by the shape of the trapping potential.
	We expect the atomic ensemble to be strongly defined in the transverse plane, and to simplify our model we assume a uniform dependence on $z$ over a finite length $L$.
	As such we write the atomic density as
	\begin{equation}
		\label{Equation: Atomic Number Density}
		n\left(\rho,\theta,z\right) = 
		\left\{  \begin{array}{c@{\qquad}l} 
		\frac{N}{2\pi \rho L} f_{P}\left(\rho\right)  & \mathrm{for}\,0 \le z \le L \\  
		0 &  \mathrm{otherwise}
		\end{array}\right.
	\end{equation}
	where $f_{P}\left(\rho\right)$ is the radial probability density function (PDF), or equivalently the radial population density, that satisfies the condition $\int_{0}^{\infty}f_{P}\left(\rho\right)\,d\rho = 1$ such that the integrated density returns the total number of atoms, $N$.
	The likelihood of finding an atom at a radius $\rho$ is determined using the Boltzmann factor by calculating the probability that a state with energy $E=U_\mathrm{d}\left(\rho\right)$ is occupied \cite{Grimm2000}.
	This process, described in detail in \ref{Appendix: Radial DPF}, results in 
	\begin{equation}
		\label{Equation: Radial PDF}
		f_{P}\left(\rho\right) = \frac{4\alpha\rho}{w^2}\exp\left(\frac{-2\alpha\rho^2}{w^2}\right),
	\end{equation}
	where we have introduced the parameter ${\alpha = -U_0/\left(k_B T\right)}$ as the magnitude of the trap depth relative to the thermal energy of the ensemble.
	This PDF represents a Gaussian density distribution centred on axis with $1/e$ radius of $w/\sqrt{2\alpha}$.
	
	We introduce a weak probe beam that is spatially matched to the dipole beam and interrogates an auxiliary upper state $\ket{\mathrm{c}}$.
	We consider this state sufficiently detuned from the dipole trap that it remains unperturbed.
	In the absence of the dipole trap, the effect of the atomic ensemble on a resonant probe field is given by
	\begin{equation}
		\label{Equation: Delta Power}
		\rmd{P_\mathrm{p}\left(\bi{r}\right)} = - \sigma n\left(\bi{r}\right) I_\mathrm{p}\left(\bi{r}\right) \rmd{V},
	\end{equation}
	where $P_\mathrm{p}$ is the power of the probe, $I_\mathrm{p}$ is the optical intensity of the probe, and $\sigma$ is the transition-specific scattering cross section of the atom.
	Rearranging and integrating in cylindrical coordinates gives us total transmission on-resonance of $\mathcal{T}=\exp\left(-\mathcal{D}_\mathrm{opt}\right)$, where optical depth is:
	\begin{eqnarray}
		\label{Equation: On Resonance OD}
		\mathcal{D}_\mathrm{opt} &=
		\int_{0}^{L} \int_{0}^{2\pi} \int_{0}^{\infty} \frac{\sigma N}{2\pi L \rho} f_{P}\left(\rho\right) \frac{2}{\pi w^2} u\left(\rho\right) \rho\, \rmd{\rho} \rmd{\theta} \rmd{z}\\
		&=  N\frac{2 \sigma}{\pi w^2} \int_{0}^{\infty} f_{P}\left(\rho\right) u\left(\rho\right) \, \rmd{\rho}\\
		&= N \frac{2\sigma}{\pi w^2} \eta
	\end{eqnarray}
	where $\eta$ is the geometrical overlap between the atomic density and optical field strength.
	For our choice of $f_{P}\left(\rho\right)$ and $u\left(\rho\right)$, we find that the geometric overlap can be given analytically as
	\begin{equation}
		\label{Equation: Geometric Overlap}
		\eta=\frac{\alpha}{1+\alpha}
	\end{equation}
	which approaches unity for a deep trap (large $\alpha$).
	
	To include the effect of the dipole trap on the system we calculate the spatially dependent light shift in $\ket{\mathrm{a}}$ that will be experienced by the probe beam.
	This relation is simply given by
	\begin{equation}
		\label{Equation: Light Shift}
		\delta_\mathrm{LS}\left(\rho\right)=-U_\mathrm{d}\left(\rho\right)/\hbar,
	\end{equation}
	which we can rearrange using knowledge of the shape of the potential to find the radial location as a function of the light shift:
	\begin{equation}
		\label{Equation: Radial Function of Light Shift}
		\rho\left(\delta_\mathrm{LS}\right)=\frac{w}{\sqrt{2}}\sqrt{\ln\left(\frac{-U_0}{\delta_\mathrm{LS} \hbar}\right)}.
	\end{equation}
	We now calculate the atomic distribution in terms of the light shift, which can be found using the following change of variables:
	\begin{eqnarray}
		\label{Equation: Change of Variables}
		f_\Delta\left(\delta_\mathrm{LS}\right)&=\left|\frac{\rmd{\rho\left(\delta_\mathrm{LS}\right)}}{\rmd{\delta_\mathrm{LS}}}\right| \cdot f_{P}\left(\rho\left(\delta_\mathrm{LS}\right)\right)\\
		&=\frac{\alpha}{\delta_\mathrm{LS}}\left(\frac{-U_0}{\delta_\mathrm{LS} \hbar}\right)^{-\alpha},
	\end{eqnarray}
	which integrates to unity over the bounds ${0<\delta_\mathrm{LS}\le -U_0/\hbar}$.
	This dipole trap intensity profile in this basis is given by $u\left(\delta_\mathrm{LS}\right)=-\delta_\mathrm{LS} \hbar/U_0$.
	
	The final piece required to calculate the light-shift perturbed spectrum is the absorption profile of the probe transition.
	Assuming the probe transition is Lorentzian in lineshape with linewidth $\Gamma_\mathrm{p}$, the absorption profile is given by
	\begin{equation}
		\label{Equation: Lorentzian Lineshape}
		\mathcal{L}\left(\Delta_\mathrm{p}\right)=\frac{1}{1+4\left(\Delta_\mathrm{p}/\Gamma_\mathrm{p}\right)^2}.
	\end{equation}
	We are now able to calculate the probe-detuning dependent transmission spectrum $\mathcal{T}\left(\Delta_\mathrm{p}\right)$ by performing the integral in \autoref{Equation: On Resonance OD} but where we also include the Lorentzian profile from \autoref{Equation: Lorentzian Lineshape} and change the integration variable to $\delta_\mathrm{LS}$.
	\begin{eqnarray}
		\label{Equation: Light Shift Optical Depth}
		\mathcal{T}\left(\Delta_\mathrm{p}\right)=
		\exp\left(-\int_{0}^{-U_0/\hbar} N \frac{2 \sigma}{\pi w^2} f_\Delta\left(\delta_\mathrm{LS}\right) u\left(\delta_\mathrm{LS}\right) \mathcal{L}\left(\Delta_\mathrm{p}-\delta_\mathrm{LS}\right) \rmd{\delta_\mathrm{LS}}\right)\nonumber\\
		\fl=\exp\left(-\frac{\mathcal{D}_\mathrm{opt}}{1+4\left(\Delta_\mathrm{p}/\Gamma_\mathrm{p}\right)^2} \frac{2}{\Gamma_\mathrm{p}} \right.\nonumber\\
		\left.\times\,\mathrm{Im}\left[\left(\Delta_\mathrm{p}+\rmi \Gamma_\mathrm{p}/2\right) {_2F_1}\left(1,\alpha+1;\alpha+2;\frac{-U_0 /\hbar}{\Delta_\mathrm{p}-\rmi \Gamma_\mathrm{p}/2}\right)\right]\right)
	\end{eqnarray}
	Here ${_2F_1}\left(a,b;c;z\right)$ is the hypergeometric function, which for physical choices of $\alpha$, is quick to evaluate, and can easily be fit to experimental data over all four physical parameters in real time.

\section{Parameter determination}
	\label{Section: Parameter Determination}
	
	Having calculated the transmission spectrum of a light-shift broadened atomic ensemble, we now seek to understand its shape as well as its dependence on the underlying physical parameters.
	To do this we calculate the spectral lineshape, and analyse physical meaning behind the mathematical parameters remaining in our model, and their effect on the resulting spectrum.
	
	\begin{figure}
		\begin{indented}
			\item[]
			\includegraphics{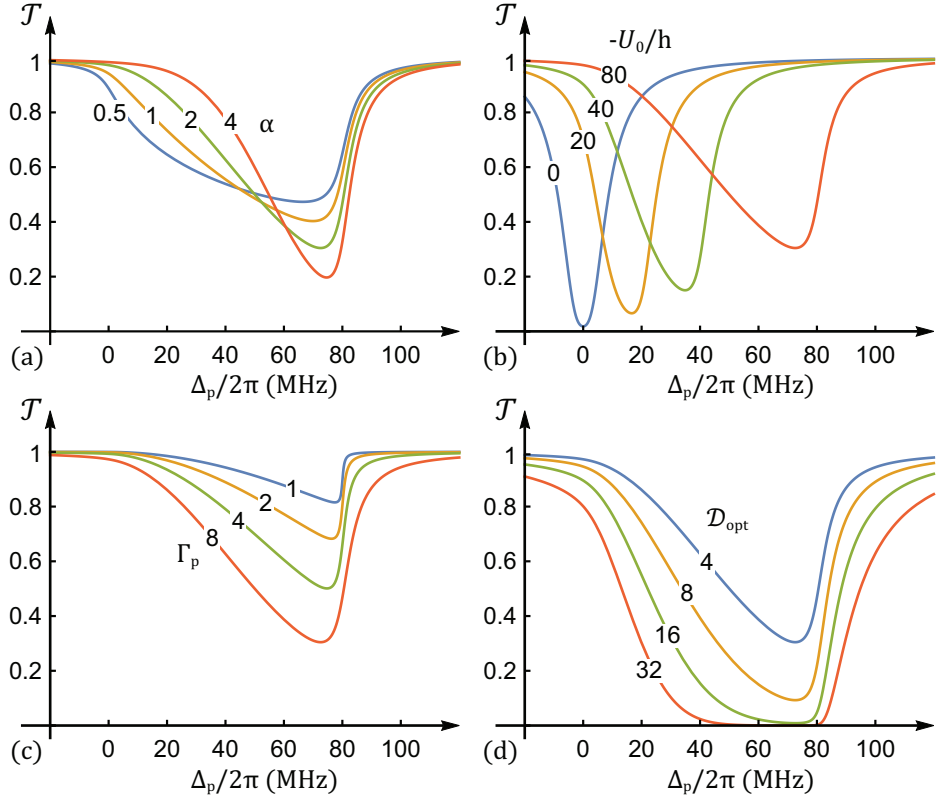}
			\caption{
				\label{Figure: Parameter Variation}
				Demonstration of the effect on the modelled transmission spectrum for variation in each physical parameter.
				In each subfigure a single parameter is varied: a) $\alpha$, b) $U_0$, c) $\Gamma_\mathrm{p}$, and d) $\mathcal{D}_\mathrm{opt}$, while the other parameters are held constant.
				In all subfigures the unvaried parameters are $\alpha=2$, $-U_0/\mathrm{h}=\SI{80}{\mega\hertz}$, $\Gamma_\mathrm{p}=2\pi\times\SI{8}{\mega\hertz}$, and $\mathcal{D}_\mathrm{opt}=4$.				
			}
		\end{indented}
	\end{figure}
	There are four physical parameters present in \autoref{Equation: Light Shift Optical Depth}: $\alpha$, the ratio of trap-depth to ensemble kinetic energy; $U_0$, the on-axis trap depth - or equivalently expressed as $-U_0/h$, the peak light-shift due to the trap; $\Gamma_\mathrm{p}$, the probe transition linewidth for atoms in the trap; and $\mathcal{D}_\mathrm{opt}$, the optical depth on the probe transition.
	All four of these parameters provide insight into the trapped ensemble that are not always derivable from conventional measurements.
	For example, the extracted value of $U_0$ from this model is an in-situ measurement of the strength of the interaction between the dipole-trap laser and the atomic ensemble that does not rely on knowledge of the optical power, the size of the beam, or the quantum state of the ensemble.
	As such this can be a powerful tool for verification of the true conditions experienced by the atoms.
	
	Similarly the measurement of $\alpha$, and thus the temperature $T$, does not rely on imaging the free expansion of the ensemble over a long relaxation time and many runs of the experiment.
	This eliminates the effect of shot-to-shot variation in the measured temperature, allowing one to instead track these processes on a nearly continuous basis.

	To understand the influence of these parameters on the absorption spectrum, we calculate it for a realistic choice of experimental parameters and systematically vary one parameter at a time over a range of values.
	We choose the common parameter set to be: $\alpha=2$, $-U_0/h=\SI{80}{\mega\hertz}$, $\Gamma_\mathrm{p}=2\pi\times \SI{8}{\mega\hertz}$, and $\mathcal{D}_\mathrm{opt}=4$.
	The calculated transmission curve for this set of values is shown in \autoref{Figure: Parameter Variation}, where each parameter is varied in subfigures (a) through (d) respectively.
	A brief description of the influence of each parameter is given below.
	
	The parameter $\alpha$ determines the breadth of the spectrum, predominantly modifying the shape of the low frequency side of the absorption peak.
	This is due to the effect $\alpha$ has on the spatial extent of the atomic distribution: tightly confined ensembles are held on axis and experience a single light shift, while weakly trapped ensembles sample a large range of the trap intensity, resulting in a wide range of light-shifts.	
	The trap-depth itself determines the depth of the potential, and as such the largest light-shift experienced.
	The dependence of $\alpha$ on $U_0$ makes these parameters interdependent, and for a fixed temperature ensemble, inversely proportional.	
	The transition linewidth strongly affects the sharpness of the high $\Delta_\mathrm{p}$ edge of the spectrum, which is physically determined by the high density of atoms that are closest to the optical axis.
	The optical depth remains as a simple scaling factor on the overall absorption of the ensemble.
	Broadening of the absorption feature is seen for high $\mathcal{D}_\mathrm{opt}$, as is typical for high dense, strongly interacting samples.

\section{Atomic distribution models}
	\label{Section: Atomic Distribution Models}
	
	Until now we have used a Gaussian distribution for the atomic density, based on the expected distribution from a well founded statistical mechanics approach.
	While we have no reason not to expect the thermodynamic derivation to well describe a system such as a laser-cooled and trapped atomic ensemble, there may be circumstance in which a non-central distribution may appear.
	There has been speculation in the literature that this might be the case for a very tightly confined ensemble within a \SI{7}{\micro\meter} core hollow optical fibre \cite{Peyronel2012}.
	
	We have used the idea as a means for testing the sensitivity of our technique to variations in atomic distribution within the trap.
	To do this, we construct a model from first principles that predicts a non-central distribution of atoms where we assume that all atoms in the trap are enforced to undergo pure circular motion in the transverse plane.
	Using the kinetic energy of the atom to determine the inwards acceleration required to maintain a constant radius, we are able to map the thermal distribution of energy onto the radial distribution of atoms within the trap, and as such find an expression for $f_{P,\,\mathrm{Ring}}\left(\rho\right)$.
	The derivation for this term is given in full in \ref{Appendix: Ring-like atomic distribution}, where the result is
	\begin{eqnarray}
		\label{Equation: Ring Distribution}
		f_{P,\,\mathrm{Ring}}\left(\rho\right)=\\
		\fl \left\{  \begin{array}{c@{\qquad}l} 
		8\sqrt{\frac{2\alpha^3}{\pi}} \frac{\rho^2}{w^3} \left|1-2\frac{\rho^2}{w^2}\right|\exp\left(-\frac{\rho^2}{w^2}\left[3+2\alpha\exp\left(-\frac{2\rho^2}{w^2}\right)\right]\right)  & \mathrm{for}\quad0 \le \rho \le w/\sqrt{2}\nonumber\\  
		0 &  \mathrm{for}\quad\rho>w/\sqrt{2}
		\end{array}\right.
	\end{eqnarray}
	which integrates to unity over $0\le\rho\le w/\sqrt{2}$ for $\alpha\gg1$.
	
\begin{figure}
		\begin{indented}
			\item[]
			\includegraphics{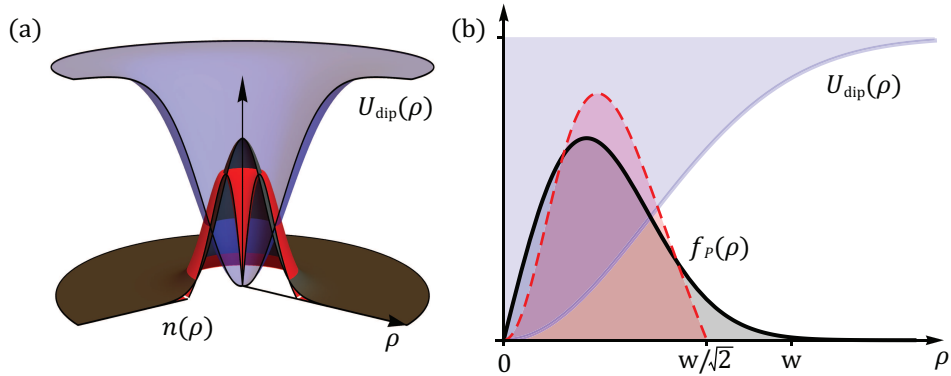}
			\caption{
				\label{Figure: Distribution Plot}
				Depiction of the Gaussian and ring-like atomic distributions used for a value of $\alpha=3$.
				a) The spatial atomic density $n\left(\rho\right)$, where the Gaussian model (Black) has a maximum density on axis, and the ring-like model (Red) is zero on axis.
				b) The radial population density $f_{P}\left(\rho\right)$ which is proportional to $\rho\times n\left(\rho\right)$, where both models have a zero on axis, but the ring-like model (Red, dashed) is also zero for $\rho>w/\sqrt{2}$.
				The dipole potential (pale blue) is shown for both cases as a visual aid.
			}
		\end{indented}
	\end{figure}

	We show both the Gaussian and ring-like atomic distributions in \autoref{Figure: Distribution Plot}.
	In (a) the spatial density $n\left(\rho\right)$ is displayed, where the Gaussian model is clearly maximum on-axis, while the ring-like model is zero on axis and peaks off-axis.
	When the radial PDF is calculated in (b), both models have peak population off-axis.
	At first sight this is surprising, however it is a consequence of the scaling of the area in an infinitesimal radial band with radius, given by $\rho\,\rmd\rho$.
	As a result, while the two models for atomic distribution are essentially orthogonal in density, in radial population density they are remarkably similar.
	The primary difference between the two models is that, as a result of the circular motion condition enforced in the derivation, the ring-like distribution is has an upper bound at $\rho=w/\sqrt{2}$.
	On the other hand the Gaussian model is able to extend indefinitely in radius, giving it a distinctly different behaviour for medium to small values of $\alpha$.  

\section{Experimental comparison}
	\label{Section: Experimental Comparison}
	We test our approach by looking at a cold-atom ensemble that has been loaded into a hollow-core optical fibre.
	The experimental setup is described in detail in \cite{Hilton2018}, however a brief overview is provided below.
	
	A magneto optical trap (MOT) prepares a sample of \num{e9} $^{85}$Rb atoms a short distance above the tip of a \SI{10}{\centi\meter} long segment of \SI{45}{\micro\meter} core kagome-lattice hollow-core photonic-crystal fibre (HC-PCF \cite{Couny2006}).
	A \SI{1}{\watt} dipole trap beam detuned $~\SI{1}{\tera\hertz}$ below the $D_1$ transition is coupled through the fibre from below, intersecting the cold-atom cloud.
	Upon release of the MOT fields, the atoms that begin within the dipole trap are confined during their fall under gravity and guided into the fibre core.
	Once inside the fibre the atoms are interrogated by a counter-propagating probe field tuned to the $F=3\rightarrow F'=4$ cycling transition on the $D_2$ line.
	This probe light is separated from the dipole trap after exiting the fibre and is incident upon an avalanche photodiode.
	
\begin{figure}
		\begin{indented}
			\item[]
			\includegraphics{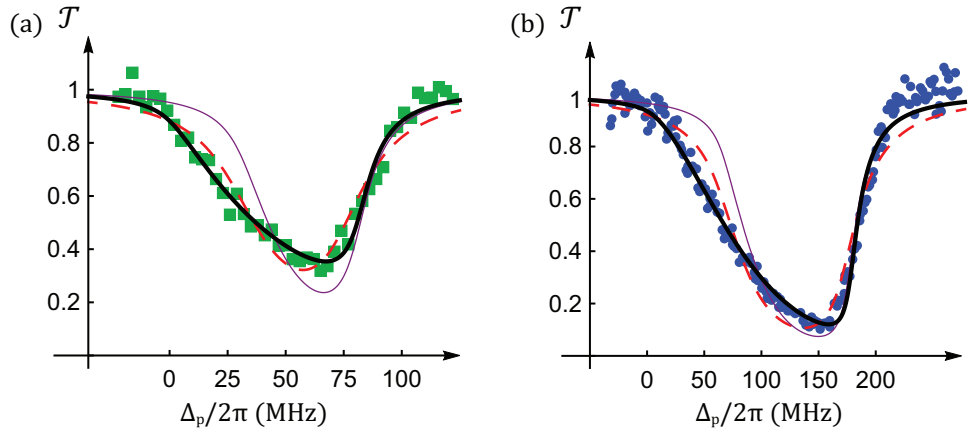}
			\caption{
				\label{Figure: Transimssion Spectra}
				Transmission measurements of an atomic ensemble in a dipole trap, with (a) our data (green squares) and (b) data from \cite{Peyronel2012} (blue circles).
				Both sets of data are shown with fitted light-shift spectra based on the Gaussian model (thick black curve), the ring-like model (dashed red curve), and the ring-like model using the fit parameters from the Gaussian (thin purple curve).
			}
		\end{indented}
	\end{figure}

	The interrogation protocol consists of a series of short pulses of weak probe light, each pulse stepped in frequency using a pair of pre-programmed waveforms that are fed to two acousto-optic modulators.
	Using this technique we are able to measure a \SI{144}{\mega\hertz} span in a single \SI{98}{\micro\second} window, allowing a true 'snapshot' of the atomic absorption to be taken.
	Conventionally the dipole trap would be switched off during probing, however here we leave it on to measure its effect on the spectrum.
	This data is shown as the green squares in \autoref{Figure: Transimssion Spectra} (a), where the spectrum has been spread over a \SI{100}{\mega\hertz} range above the unperturbed linecentre.
	
	To compare our models to this data we run a least-squares fitting algorithm where all four physical parameters listed above are allowed to be free.
	We show the outcome of the fit for both Gaussian (black, solid) and also for a possible ring-like (red, dashed) atomic distributions.
	It is clear that the closest fit to our results is given by the fitted Gaussian model, which captures the full asymmetry seen in the experimental data.
	Although the ring-like distribution can match the depth, location, and width of the absorption feature, it fails to capture the asymmetric shape.
	The best fit parameters with associated uncertainties are given in \autoref{Table: Transmission Fit Results}, where the ensemble temperature is calculated for the Gaussian model fits using  $T=-U_0 / \alpha k_B$.
	Importantly the parameters returned by the ring-like distribution fit are obviously unphysical and thus we deduce that this solution is unlikely to explain the results.
	In the case where we copy physically plausible values for the parameters from the Gaussian distribution fit into the ring-like solution (thin, purple curve in \autoref{Figure: Transimssion Spectra}) it is clear that the it still possesses the wrong shape to explain our observations.
		
	In addition to our experimental results, we perform the same analysis to the data used by \emph{Peyronel et al.} that the authors indicated had suggested a ring-like distribution, in figure 6(A.1) in \cite{Peyronel2012}.
	We show this data again in \autoref{Figure: Transimssion Spectra} (b) as blue circles, with the theory curves following the same pattern as before.
	As with our work, the best fit is found by the Gaussian distribution which fully captures the shape of the data, while both ring-like models are unable to capture the asymmetry in the light-shift broadened spectrum.
	
	\begin{table}
		\caption{
			\label{Table: Transmission Fit Results}
			Results of fitting to AC-Stark shifted data with mathematical models based on ring-like and Gaussian atomic distributions.
		}
		\begin{indented}
			\item[]
			\begin{tabular}{@{}l  l   S[table-format=1.2(1)] S[table-format=1.1(1)] S[table-format=3.0(1)]   S[table-format=1.1(1)]   S[table-format=2.0(1)]  }
				\br
				& & & \multicolumn{1}{c}{$T$} &\multicolumn{1}{c}{ $-U_0 /h$} & & \multicolumn{1}{c}{$ \Gamma_\mathrm{p}/2\pi$}\\
				Source & Model & \multicolumn{1}{c}{$\alpha$}  & \multicolumn{1}{c}{(\SI{}{\milli\kelvin})} & \multicolumn{1}{c}{(\SI{}{\mega\hertz})} & \multicolumn{1}{c}{$\mathcal{D}_\mathrm{opt}$} & \multicolumn{1}{c}{(\SI{}{\mega\hertz})} \\
				\mr
				\multirow{2}{*}{Our Data} & Gaussian & 1.0 \pm 0.1 & 4.0 \pm 0.4 & 81 \pm 1 & 2.8\pm 0.3 & 16 \pm 2 \\ 
				& Ring-like  & 0.0 \pm 4 & & 74 \pm 8 & 1.5 \pm 0.2 & 32 \pm 6 \\ 
				\mr
				\multirow{2}{*}{\emph{Peyronel et al.}} & Gaussian & 1.59 \pm 0.07 & 5.4 \pm 0.2 & 177 \pm 1 & 8.0 \pm 0.6 & 17 \pm 1 \\
				 & Ring-like & 0.0 \pm 2 & & 172 \pm 5 & 4.4 \pm 0.4 & 33 \pm 2 \\ 
				 \br
			\end{tabular} 
		\end{indented}
	\end{table}

	It is worth mentioning that the linewidth $\Gamma_\mathrm{p}/2\pi$ is consistently much larger than the expected natural linewidth of $\sim$\SI{6}{\mega\hertz}.
	Assuming that the atoms are otherwise unperturbed, this additional broadening can be caused by two effects: power broadening by spontaneous absorption of the trap beam or differential light shifts associated with different $m_F$ ground states.
	Power broadening is unlikely due to the weak scattering of trap photons which, for our experiment is calculated to be less than $\SI{10}{\kilo\hertz}$.
	On the other hand, differential light-shifts arise from variance in the Clebsch-Gordan coefficients, and hence coupling strength, between the $m_F \rightarrow m_F'$ manifold in the trapping transition. 
	As a result, atoms in the $2n+1$  Zeeman sub-states experience different trap depths, the overlap of which produces a broader spectrum.
	A more appropriate fit function would be the product of $2n+1$ separate transmission spectra with $U_0$, $\alpha$, and $\mathcal{D}_\mathrm{opt}$ for each spectrum appropriately weighted by the relative transition strengths.
	
\section{Conclusion}
	\label{Section: Conclusion}
	
	We have developed a technique that employs the spatially varying light-shift inherent to a dipole trap as a means to perform spectroscopy on the trapped atomic ensemble.
	The trap itself provides the perfect reference, mapping the radial location of each atom into a unique frequency shift.
	Using an understanding of the shape of the dipole trap and the resulting spatial distribution of atoms, we produce a testable model of the light-shift broadened atomic absorption spectrum.
	This technique is able to rapidly infer the number of atoms, temperature of the ensemble, depth of trap, and transition linewidth, with a high level of independence between each of the experimental parameters.
	
	We experimentally test this technique using a hollow-core fibre loaded cold-atom ensemble and are able to take a single-shot snapshot of the light-shift broadened spectrum.
	Fitting to this data with two models for the distribution of atoms in a radial trapping field, we find strong agreement with the Gaussian distribution model.
	This comparison additionally demonstrates that our spectroscopic technique is sensitive to the use of an appropriate model for the atomic distribution.
	Using the fit parameters we able to extract relevant physical properties of the atomic ensemble and dipole trap from the shape of the measured spectrum.
	
	We hope this analysis provides insight into the dynamics of a trapped atomic system, and expect our interrogation scheme to be instrumental in acquiring rapid feedback on the parameters of state of cold atom systems that would otherwise require slow and repetitive interrogation sequences over many experimental cycles.
	
	\ack
	We would like to thank the South Australian government for supporting this research through the PRIF program.
	
	This research was funded by the Australian government through the Australian Research Council (Project: DE12012028).
	
	We would also like to thank \textit{Peyronel et al.} for providing their data for use as a comparison.
	\appendix
	\section{Calculation of radial probability density function}
		\label{Appendix: Radial DPF}
		To find the likelihood of finding an atom at a radius $\rho$, we make use of fundamental thermodynamic arguments.
		The Boltzmann factor predicts the probability of a state with energy $E=U_\mathrm{d}\left(\bi{r}\right)$ being occupied.
		The radially dependent trap depth defines the energy of the state, allowing us to find that the atomic density is of the form
		\begin{eqnarray}
			\label{Equation: Gaussian Number Distribution}
			n\left(\rho\right) &\propto \exp\left(-U_\mathrm{d}\left(\rho\right)/k_B T\right)\nonumber\\
			&\propto \exp\left(\frac{U_0}{k_B T}\,\exp\left(-2\rho^2/w^2\right)\right)
		\end{eqnarray}
		where $k_B$ is the Boltzmann constant, and $T$ is the ensemble temperature.
		
		This is simplified to a useful form by assuming that the atoms remain close to the central axis, and as such the potential can be assumed to be harmonic.
		Taking a power series to second order in the radius, we find a Gaussian distribution centred on axis with the form
		\begin{equation}
			\label{Equation: Approximate Gaussian Number Distribution}
			n\left(\rho\right) \propto \exp\left(\alpha-\frac{-2\alpha\rho^2}{w^2}\right),
		\end{equation}
		where we have introduced the parameter $\alpha = -U_0/\left(k_B T\right)$ as the magnitude of the trap depth relative to the thermal energy of the ensemble.
		
		Integrating $n\left(\bi{r}\right)$ and using \autoref{Equation: Atomic Number Density} we calculate the normalised PDF
		\begin{equation}
			\label{Equation: Radial PDF (Appendix)}
			f_{P}\left(\rho\right) = \frac{4\alpha\rho}{w^2}\exp\left(\frac{-2\alpha\rho^2}{w^2}\right)
		\end{equation}
		which integrates to unity over $0\le\rho<\infty$.
		
\section{Ring-like atomic distribution}
	\label{Appendix: Ring-like atomic distribution}
	
	To generate a ring-like atomic distribution we begin by simplifying the atomic motion by considering only trajectories that have constant radius.
	This is helpful in two ways: it allows us to include the full Gaussian form of the potential and removes the need to perform spatial integration over the atomic trajectory for all choices of ellipticity.	
	
	The equation of motion for a circular orbit is trivial, and is given by 
	\begin{equation}
		\label{Equation: Circular Motion}
		-\bi{a}\cdot\brho = v_\perp^2
	\end{equation}
	where $v_\perp$ is the speed perpendicular to $\brho$, which is simply the magnitude of the total velocity in the transverse plane due to our choice of trajectory.
	Similarly, $\bi{a}=a_\rho \hat{\brho}$ is the radial acceleration due to the trap potential, which for a particle with mass $m$ can be calculated as a function of radius as
	\begin{eqnarray}
		\label{Equation: Acceleration In Trap}
		a_\rho\left(\rho\right) &=\frac{1}{m}\cdot\frac{-\rmd U_\mathrm{d}\left(\rho\right)}{\rmd \rho}\nonumber\\
		&=\frac{4\rho U_0}{m w^2}\exp\left(\frac{-2\rho^2}{w^2}\right).
	\end{eqnarray}
	
	We calculate the kinetic energy of an atom in this trajectory, $\epsilon$, using 
	\begin{equation}
		\label{Equation: Energy of Orbit}
		\epsilon\left(\rho\right)=\frac{1}{2}mv^2=\frac{-2\rho^2U_0}{w^2}\exp\left(\frac{-2\rho^2}{w^2}\right)
	\end{equation}
	in which we have assume the axial component is negligible.
	Assuming the atomic ensemble is in thermal equilibrium, we can describe the energy distribution in the system using a Maxwell-Boltzmann distribution:
	\begin{equation}
		\label{Equation: Boltzmann Distribution}
		f_E\left(\epsilon\right)=2\sqrt{\frac{\epsilon}{\pi}}\left(\frac{1}{k_B T}\right)^{3/2}\exp\left(\frac{-\epsilon}{k_B T}\right).
	\end{equation}
	By implementing a change of variables as in \autoref{Equation: Change of Variables}, we can generate the radial distribution of atoms for the ring-like model using \autoref{Equation: Energy of Orbit}
	\begin{eqnarray}
		\label{Equation: Ring-like Radial Distribution}
		f_{P,\,\mathrm{Ring}}\left(\rho\right) =\left|\frac{\rmd\epsilon\left(\rho\right)}{\rmd\rho}\right|\cdot f_E\left(\epsilon\left(\rho\right)\right)\nonumber\\
		= 8\sqrt{\frac{2\alpha^3}{\pi}} \frac{\rho^2}{w^3} \left|1-2\frac{\rho^2}{w^2}\right|\exp\left(-\frac{\rho^2}{w^2}\left[3+2\alpha\exp\left(-\frac{2\rho^2}{w^2}\right)\right]\right)
	\end{eqnarray}
	where again we have introduced the substitution $\alpha=-U_0/k_B T$.
	
	It should be noted that for a Gaussian beam the trap depth $U_\mathrm{d}\left(\rho\right)$ is deepest at $\rho=0$ and monotonic in $\rho$, while the orbital energy $\epsilon\left(\rho\right)$ is zero at $\rho=0$, and has a maximum at $\rho=w/\sqrt{2}$.
	From this we can infer that we expect to find two distinct circular orbits for each possible value of the kinetic energy: one for $0\le\rho<w/\sqrt{2}$, and one for $\rho>w/\sqrt{2}$.
	The set of solutions with large $\rho$ represent trajectories with orbital energies that are greater than the trap depth.
	While these are still valid solutions to the equations of motion, the orbits are unstable and an infinitesimal increase in velocity will free the atoms from the trap.
	To produce a physically realistic model we truncate the radial distribution at $\rho=w/\sqrt{2}$, and choose only to count stable orbits.
	
	While the ring-like model radial distribution $f_{P,\,\mathrm{Ring}}\left(\rho\right)$ integrates to unity over the bounds $0\le\rho\le w/\sqrt{2}$ for large $\alpha$, for small $\alpha$ this is not the case.
	In our model there is a maximum orbital energy that can maintain a circular orbit.
	For small $\alpha$ there is a non-zero fraction of the Maxwell-Boltzmann distribution that that has energy greater than this upper bound.
	Our use of the change of variables intrinsically includes this overlap issue, and the resulting integral $\int_{0}^{w/\sqrt{2}} f_{P,\,\mathrm{Ring}}\left(\rho\right)\rmd\rho$ produces the fraction of the energy distribution that is capable of being trapped.
	
	While this is useful information, in our model we have scaled the radial distribution so that it is normalized to the total atomic number in the thermal cloud that is trapped.
	\newpage	
	\section*{References}
	\bibliographystyle{unsrt}
	\bibliography{LSLpaper}

\begin{thebibliography}{10}

\bibitem{Kasevich1991}
Mark Kasevich and Steven Chu.
\newblock {Atomic interferometry using stimulated Raman transitions}.
\newblock {\em Physical Review Letters}, 67(2):181--184, jul 1991.

\bibitem{Peters2001}
A~Peters, K~Y Chung, and S~Chu.
\newblock {High-precision gravity measurements using atom interferometry}.
\newblock {\em Metrologia}, 38(1):25--61, feb 2001.

\bibitem{Canuel2006}
B.~Canuel, F.~Leduc, D.~Holleville, A.~Gauguet, J.~Fils, A.~Virdis, A.~Clairon,
  N.~Dimarcq, Ch~J. Bord{\'{e}}, A.~Landragin, and P.~Bouyer.
\newblock {Six-axis inertial sensor using cold-atom interferometry}.
\newblock {\em Physical Review Letters}, 97(1):1--4, 2006.

\bibitem{Cronin2009}
Alexander~D. Cronin, J{\"{o}}rg Schmiedmayer, and David~E. Pritchard.
\newblock {Optics and interferometry with atoms and molecules}.
\newblock {\em Reviews of Modern Physics}, 81(3):1051--1129, jul 2009.

\bibitem{Stockton2011}
J.~K. Stockton, K.~Takase, and M.~A. Kasevich.
\newblock {Absolute Geodetic Rotation Measurement Using Atom Interferometry}.
\newblock {\em Physical Review Letters}, 107(13):133001, sep 2011.

\bibitem{Hinkley2013}
N.~Hinkley, J.~A. Sherman, N.~B. Phillips, M.~Schioppo, N.~D. Lemke, K.~Beloy,
  M.~Pizzocaro, C.~W. Oates, and A.~D. Ludlow.
\newblock {An Atomic Clock with 10-18 Instability}.
\newblock {\em Science}, 341(6151):1215--1218, sep 2013.

\bibitem{Altin2013}
P.~A. Altin, M.~T. Johnsson, V.~Negnevitsky, G.~R. Dennis, R.~P. Anderson,
  J.~E. Debs, S.~S. Szigeti, K.~S. Hardman, S.~Bennetts, G.~D. McDonald, L.~D.
  Turner, J.~D. Close, and N.~P. Robins.
\newblock {Precision atomic gravimeter based on Bragg diffraction}.
\newblock {\em New Journal of Physics}, 15(2):023009, feb 2013.

\bibitem{Dutta2016}
I.~Dutta, D.~Savoie, B.~Fang, B.~Venon, C.~L. {Garrido Alzar}, R.~Geiger, and
  A.~Landragin.
\newblock {Continuous Cold-Atom Inertial Sensor with 1 nrad / sec Rotation
  Stability}.
\newblock {\em Physical Review Letters}, 116(18):183003, may 2016.

\bibitem{Sparkes2013}
B.~M. Sparkes, J.~Bernu, M.~Hosseini, J.~Geng, Q.~Glorieux, P.~A. Altin, P.~K.
  Lam, N.~P. Robins, and B.~C. Buchler.
\newblock {Gradient echo memory in an ultra-high optical depth cold atomic
  ensemble}.
\newblock {\em New Journal of Physics}, 15(8):085027, aug 2013.

\bibitem{Liu2016}
Zi-Yu Liu, Yi-Hsin Chen, Yen-Chun Chen, Hsiang-Yu Lo, Pin-Ju Tsai, Ite~A. Yu,
  Ying-Cheng Chen, and Yong-Fan Chen.
\newblock {Large Cross-Phase Modulations at the Few-Photon Level}.
\newblock {\em Physical Review Letters}, 117(20):203601, nov 2016.

\bibitem{DeAlmeida2016}
A.~J.~F. de~Almeida, M.-A. Maynard, C~Banerjee, D~Felinto, F~Goldfarb, and
  J~W~R Tabosa.
\newblock {Nonvolatile optical memory via recoil-induced resonance in a pure
  two-level system}.
\newblock {\em Physical Review A}, 94(6):063834, dec 2016.

\bibitem{Park2018}
Kwang-Kyoon Park, Young-Wook Cho, Young-Tak Chough, and Yoon-Ho Kim.
\newblock {Experimental Demonstration of Quantum Stationary Light Pulses in an
  Atomic Ensemble}.
\newblock {\em Physical Review X}, 8(2):021016, 2018.

\bibitem{Hsiao2018}
Ya-Fen Hsiao, Pin-Ju Tsai, Hung-Shiue Chen, Sheng-Xiang Lin, Chih-Chiao Hung,
  Chih-Hsi Lee, Yi-Hsin Chen, Yong-Fan Chen, Ite~A Yu, and Ying-Cheng Chen.
\newblock {Highly Efficient Coherent Optical Memory Based on
  Electromagnetically Induced Transparency}.
\newblock {\em Physical Review Letters}, 120(18):183602, may 2018.

\bibitem{Katori2003}
Hidetoshi Katori, Masao Takamoto, V.~G. Pal'chikov, and V.~D. Ovsiannikov.
\newblock {Ultrastable Optical Clock with Neutral Atoms in an Engineered Light
  Shift Trap}.
\newblock {\em Physical Review Letters}, 91(17):173005, oct 2003.

\bibitem{Takamoto2005}
Masao Takamoto, Feng-Lei Hong, Ryoichi Higashi, and Hidetoshi Katori.
\newblock {An optical lattice clock}.
\newblock {\em Nature}, 435(7040):321--324, may 2005.

\bibitem{Ye2008}
J~Ye, H~J Kimble, and H~Katori.
\newblock {Quantum State Engineering and Precision Metrology Using
  State-Insensitive Light Traps}.
\newblock {\em Science}, 320(5884):1734--1738, jun 2008.

\bibitem{Bloom2014}
B.~J. Bloom, T.~L. Nicholson, J.~R. Williams, S.~L. Campbell, M.~Bishof,
  X.~Zhang, W.~Zhang, S.~L. Bromley, and J.~Ye.
\newblock {An optical lattice clock with accuracy and stability at the 10−18
  level}.
\newblock {\em Nature}, 506(7486):71--75, feb 2014.

\bibitem{Huang2014}
J.~Huang, S.~Wu, H.~Zhong, and C.~Lee.
\newblock {\em QUANTUM METROLOGY WITH COLD ATOMS}, chapter~7, pages 365--415.
\newblock World Scientific, 2014.

\bibitem{Zhang2016}
Xibo Zhang and Jun Ye.
\newblock {Precision measurement and frequency metrology with ultracold atoms}.
\newblock {\em National Science Review}, 3(2):189--200, 2016.

\bibitem{LeTargat2013}
R.~{Le Targat}, L.~Lorini, Y.~{Le Coq}, M.~Zawada, J.~Gu{\'{e}}na, M.~Abgrall,
  M.~Gurov, P.~Rosenbusch, D.~G. Rovera, B.~Nag{\'{o}}rny, R.~Gartman, P.~G.
  Westergaard, M.~E. Tobar, M.~Lours, G.~Santarelli, A.~Clairon, S.~Bize,
  P.~Laurent, P.~Lemonde, and J.~Lodewyck.
\newblock {Experimental realization of an optical second with strontium lattice
  clocks}.
\newblock {\em Nature Communications}, 4(1):2109, dec 2013.

\bibitem{Gross2017}
Christian Gross and Immanuel Bloch.
\newblock {Quantum simulations with ultracold atoms in optical lattices}.
\newblock {\em Science}, 357(6355):995--1001, sep 2017.

\bibitem{Vorrath2010}
S.~Vorrath, S.~A. M{\"{o}}ller, P.~Windpassinger, K.~Bongs, and K.~Sengstock.
\newblock {Efficient guiding of cold atoms through a photonic band gap fiber}.
\newblock {\em New Journal of Physics}, 12:123015, 2010.

\bibitem{Bajcsy2011}
M.~Bajcsy, S.~Hofferberth, T.~Peyronel, V.~Balic, Q.~Liang, a.~S. Zibrov,
  V.~Vuletic, and M.~D. Lukin.
\newblock {Laser-cooled atoms inside a hollow-core photonic-crystal fiber}.
\newblock {\em Physical Review A - Atomic, Molecular, and Optical Physics},
  83(6):1--9, 2011.

\bibitem{Okaba2014}
Shoichi Okaba, Tetsushi Takano, Fetah Benabid, Tom Bradley, Luca Vincetti,
  Zakhar Maizelis, Valery Yampol'skii, Franco Nori, and Hidetoshi Katori.
\newblock {Lamb-Dicke spectroscopy of atoms in a hollow-core photonic crystal
  fibre}.
\newblock {\em Nature Communications}, 5:4096, jun 2014.

\bibitem{Blatt2016}
Frank Blatt, Lachezar~S. Simeonov, Thomas Halfmann, and Thorsten Peters.
\newblock {Stationary light pulses and narrowband light storage in a
  laser-cooled ensemble loaded into a hollow-core fiber}.
\newblock {\em Physical Review A}, 94(4):043833, oct 2016.

\bibitem{Xin2017}
Mingjie Xin, Wui~Seng Leong, Zilong Chen, and Shau-Yu Lan.
\newblock {An atom interferometer inside a hollow-core photonic crystal fiber}.
\newblock {\em Science Advances}, 4(1):e1701723, jan 2018.

\bibitem{Hilton2018}
A.P. Hilton, C.~Perrella, F.~Benabid, B.M. Sparkes, A.N. Luiten, and P.S.
  Light.
\newblock {High-efficiency cold-atom transport into a waveguide trap}.
\newblock {\em Physical Review Applied}, 10(4):044034, 2018.

\bibitem{Langbecker2018}
Maria Langbecker, Ronja Wirtz, Fabian Knoch, Mohammad Noaman, Thomas Speck, and
  Patrick Windpassinger.
\newblock {Highly controlled optical transport of cold atoms into a hollow-core
  fiber}.
\newblock {\em New Journal of Physics}, 20(8):083038, aug 2018.

\bibitem{Yoon2019}
Taehyun Yoon and Michal Bajcsy.
\newblock {Laser-cooled cesium atoms confined with a magic-wavelength dipole
  trap inside a hollow-core photonic-bandgap fiber}.
\newblock {\em Physical Review A}, 99(2):023415, feb 2019.

\bibitem{Chikkatur2002}
A~P Chikkatur.
\newblock {A Continuous Source of Bose-Einstein Condensed Atoms}.
\newblock {\em Science}, 296(5576):2193--2195, jun 2002.

\bibitem{Leanhardt2002a}
A.~E. Leanhardt, A.~P. Chikkatur, D.~Kielpinski, Y.~Shin, T.~L. Gustavson,
  W.~Ketterle, and D.~E. Pritchard.
\newblock {Propagation of Bose-Einstein Condensates in a Magnetic Waveguide}.
\newblock {\em Physical Review Letters}, 89(4):040401, jul 2002.

\bibitem{Leanhardt2002b}
A.~E. Leanhardt, A.~G{\"{o}}rlitz, A.~P. Chikkatur, D.~Kielpinski, Y.~Shin,
  D.~E. Pritchard, and W.~Ketterle.
\newblock {Imprinting Vortices in a Bose-Einstein Condensate using Topological
  Phases}.
\newblock {\em Physical Review Letters}, 89(19):190403, oct 2002.

\bibitem{Leanhardt2003}
A~E Leanhardt.
\newblock {Cooling Bose-Einstein Condensates Below 500 Picokelvin}.
\newblock {\em Science}, 301(5639):1513--1515, sep 2003.

\bibitem{Ketterle1999}
W.~Ketterle, D.~S. Durfee, and D.~M. Stamper-Kurn.
\newblock {Making, probing and understanding Bose-Einstein condensates}.
\newblock In {\em Proceedings of the International School of Physics "Enrico
  Fermi"}, volume 140, pages 67--176, 1999.

\bibitem{Peyronel2012}
Thibault Peyronel, Michal Bajcsy, Sebastian Hofferberth, Vlatko Balic, Mohammad
  Hafezi, Qiyu Liang, Alexander Zibrov, Vladan Vuletic, and Mikhail~D. Lukin.
\newblock {Switching and counting with atomic vapors in photonic-crystal
  fibers}.
\newblock {\em IEEE Journal on Selected Topics in Quantum Electronics},
  18(6):1747--1753, 2012.

\bibitem{Blatt2014}
F.~Blatt, T.~Halfmann, and T.~Peters.
\newblock {One-dimensional ultracold medium of extreme optical depth}.
\newblock {\em Optics Letters}, 39(3):446, 2014.

\bibitem{Haffner2003}
H.~H{\"{a}}ffner, S.~Gulde, M.~Riebe, G.~Lancaster, C.~Becher, J.~Eschner,
  F.~Schmidt-Kaler, and R.~Blatt.
\newblock {Precision Measurement and Compensation of Optical Stark Shifts for
  an Ion-Trap Quantum Processor}.
\newblock {\em Physical Review Letters}, 90(14):143602, apr 2003.

\bibitem{Hong2005}
Tao Hong, Claire Cramer, Warren Nagourney, and E.~N. Fortson.
\newblock {Optical Clocks Based on Ultranarrow Three-Photon Resonances in
  Alkaline Earth Atoms}.
\newblock {\em Physical Review Letters}, 94(5):050801, feb 2005.

\bibitem{Santra2005}
Robin Santra, Ennio Arimondo, Tetsuya Ido, Chris~H. Greene, and Jun Ye.
\newblock {High-Accuracy Optical Clock via Three-Level Coherence in Neutral
  Bosonic $^{88}$Sr}.
\newblock {\em Physical Review Letters}, 94(17):173002, may 2005.

\bibitem{Grimm2000}
Rudolf Grimm, Matthias Weidem{\"{u}}ller, and Yurii~B. Ovchinnikov.
\newblock {Optical Dipole Traps for Neutral Atoms}.
\newblock {\em Adv. At. Mol. Opt. Phys.}, 42:95--170, 2000.

\bibitem{Couny2006}
F~Couny, Fetah Benabid, and P. S. Light.
\newblock {Large-pitch kagome-structured hollow-core photonic crystal fiber}.
\newblock {\em Optics Letters}, 31(24):3574--3576, 2006.

\end{thebibliography}
\end{document}